# Heavy Element Abundances in Presolar Silicon Carbide Grains from Low-Metallicity AGB Stars

*Peter Hoppe[A,E], Jan Leitner[A], Christian Vollmer[A], Elmar Gröner[A], Philipp R. Heck[A,B], Roberto Gallino[C], and Sachiko Amari[D]*

[A]Max-Planck-Institute for Chemistry, P.O. Box 3060, D-55020 Mainz, Germany.
[B]Chicago Center for Cosmochemistry, Dept. of the Geophysical Sciences, The University of Chicago, Chicago, IL 60637, USA.
[C]Dipartimento di Fisica Generale, Università di Torino, 10125 Torino, Italy.
[D]Washington University, Laboratory for Space Sciences & the Physics Department, St. Louis, MO 63130, USA.
[E]E-mail: hoppe@mpch-mainz.mpg.de

**Abstract:** Primitive meteorites contain small amounts of presolar minerals that formed in the winds of evolved stars or in the ejecta of stellar explosions. Silicon carbide is the best studied presolar mineral. Based on its isotopic compositions it was divided into distinct populations that have different origins: Most abundant are the mainstream grains which are believed to come from 1.5-3 $M_\odot$ AGB stars of roughly solar metallicitiy. The rare Y and Z grains are likely to come from 1.5-3 $M_\odot$ AGB stars as well, but with subsolar metallicities (0.3-0.5x solar). Here we report on C and Si isotope and trace element (Zr, Ba) studies of individual, submicrometer-sized SiC grains. The most striking results are: (1) Zr and Ba concentrations are higher in Y and Z grains than in mainstream grains, with enrichments relative to Si and solar of up to 70x (Zr) and 170x (Ba), respectively. (2) For the Y and Z grains there is a positive correlation between Ba concentrations and amount of s-process Si. This correlation is well explained by predictions for 2-3 $M_\odot$ AGB stars with metallicities of 0.3-0.5x solar. This confirms low-metallicity stars as most likely stellar sources for the Y and Z grains.

## 1. Introduction

Primitive meteorites and interplanetary dust particles contain small amounts (at the level of ppb to hundreds of ppm) of refractory dust grains that formed in the winds of evolved stars or in the ejecta of stellar explosions (for recent reviews see: Lodders & Amari 2005; Zinner 2007; Hoppe 2008). These "presolar" grains are characterized by large isotopic anomalies in the major and minor/trace elements. Among the identified presolar minerals are diamond, silicon carbide (SiC), graphite, silicon nitride ($Si_3N_4$), corundum and other forms of $Al_2O_3$, spinel ($MgAl_2O_4$), hibonite ($CaAl_{12}O_{19}$), titanium oxide, wüstite (FeO), and silicates (amorphous and crystalline). Laboratory studies of those grains have provided a wealth of information on stellar nucleosynthesis and evolution, mixing in stellar ejecta, the Galactic chemical evolution, dust formation in circumstellar environments, chemical and physical processes in the interstellar medium, and the inventory of stars that contributed dust to the molecular cloud from which our Solar System formed some 4.6 Gy ago.

Silicon carbide is the best studied presolar mineral phase. Based on the isotopic compositions of C, N, and Si it was divided into distinct populations (Hoppe & Ott 1997): The mainstream grains, which comprise about 90% of all meteoritc SiC, the minor type AB, X, Y, and Z grains, and the very rare nova grains. Asymptotic giant branch (AGB) stars are the most likely stellar sources for the vast majority of the SiC grains. This has been inferred from a comparison of the isotope data of the grains with stellar models. The mainstream grains apparently come from 1.5-3 $M_\odot$ AGB stars of roughly solar metallicity (Lugaro et al. 2003). Also the Y and Z grains are believed to come from such stars, but with sub-solar metallicities (0.3-0.5x solar) (Amari et al. 2001; Hoppe et al. 1997; Nittler & Alexander 2003; Zinner et al. 2007; Zinner et al. 2006). The Y and Z grains exhibit specific Si-isotopic signatures, namely, a shift to $^{30}$Si-rich compositions compared to the Si-isotopic signature of the mainstream grains and, for the Z grains, lower than solar $^{29}$Si/$^{28}$Si. The shift has been explained by dredge-up of Si, whose isotopic composition was altered by neutron-capture reactions (in the following referred to as "s-process Si") in their parent stars. While Y grains have $^{12}$C/$^{13}$C > 100, Z grains have $^{12}$C/$^{13}$C < 100. The comparatively low $^{12}$C/$^{13}$C ratios in Z grains have been taken as evidence for the operation of cool bottom processing (Wasserburg, Boothroyd, & Sackmann 1995; Nollett, Busso, & Wasserburg 2003) in the parent stars of Z grains (Hoppe et al. 1997; Nittler & Alexander 2003; Zinner et al. 2006).

Previous measurements (e.g., Zinner et al. 2007) showed that the abundance of Z grains is much higher among submicrometer-sized grains than among micrometer-sized grains for which most of the isotope data exist. In this paper, we report on an extended search for type Y and Z grains in the Murchsion SiC separate KJB (typical size 0.25-0.45 μm; Amari, Lewis, & Anders 1994) and on trace element studies (Zr, Mo, Ba, Nd) of selected Y and Z grains in order to get further insights into their origins.

## 2. Experimental

Thousands of submicrometer-sized KJB grains were dispersed on an ultra-clean gold foil in an 4:1 isopropanol-water suspension. Carbon and Si isotope measurements were done by Secondary Ion Mass Spectrometry (SIMS) in the fully automated ion imaging procedure developed for the NanoSIMS ion probe at the Max-Planck-Institute for Chemistry (Gröner & Hoppe 2006). This procedure consists of three steps: (1) Acquisition of simultaneous $^{12}$C$^-$, $^{13}$C$^-$, $^{28}$Si$^-$, $^{29}$Si$^-$, and $^{30}$Si$^-$ ion images, produced by rastering a focussed Cs$^+$ primary ion beam (100 nm, ~1 pA) over areas 30 x 30 μm$^2$ in size (total integration time of ~15 minutes). (2) Automated particle recognition and C and Si isotope measurements in square areas around each grain (with lateral length of 2x the grain size, defined at 10% of the maximum $^{28}$Si intensity, and with integration time of 60 s). (3) Moving the sample stage to the adjacent 30 x 30 μm$^2$-sized analysis area and repetition with step (1). Application to 1 μm-sized synthetic SiC grains resulted in grain-to-grain reproducibilities for C and Si isotope ratios of better than 10 ‰, which is by far sufficient for isotope studies of presolar SiC grains.

Selected Y and Z grains as well as several mainstream grains were subsequently measured for Zr, Mo, Ba, and Nd concentrations. For this purpose a focussed primary O$^-$ ion beam (300 nm, ~10-15 pA) was rastered over the grains (2 x 2 μm$^2$) and positive secondary ions of $^{28}$Si, $^{90}$Zr, $^{98}$Mo, $^{138}$Ba, and $^{144}$Nd were measured in multi-collection. For the quantification of trace element concentrations, relative SIMS sensitivity factors were measured on fine-grained NIST SRM 611 silicate glass fragments dispersed on a gold foil.

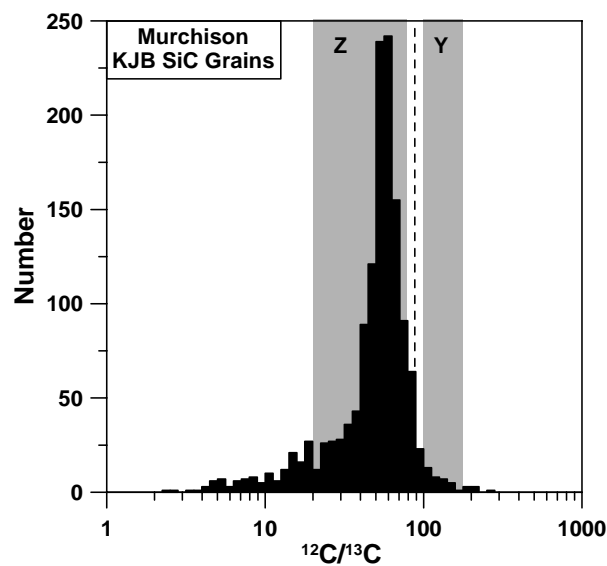

**Figure 1** Histogram of $^{12}$C/$^{13}$C ratios of presolar SiC grains from Murchison separate KJB. The ranges observed for Y and Z grains are indicated by the grey-shaded areas.

## 3. Results and Discussion

The automated ion imaging identified some 1300 SiC grains on the KJB mount. Among these grains are 88% mainstream grains, which is compatible with what was observed for micrometer-sized grains (Hoppe et al. 1994; Nittler & Alexander 2003). This also holds for the abundances of the rare type AB, Y, and X grains. The situation is clearly different for the type Z grains which make up 5.4% of the analyzed KJB grains. This is much higher than observed for micrometer-sized grains but is compatible with the results obtained for submicrometer-sized SiC grains from the Indarch enstatite chondrite (Zinner et al. 2007). The predominant presence of Z grains among smaller SiC grains may be explained by relative low Si abundances in the winds of low-metallicity AGB stars that prevents the growth of larger SiC particles (Gail et al. 2008).

Mass-weighted averages of C- and Si-isotopic compositions are 38.3 ± 1.1 ($^{13}C/^{12}C$), 25.9 ± 1.4 ‰ ($\delta^{29}Si$), and 37.2 ± 1.9 ‰ ($\delta^{30}Si$). This is in agreement with the data for KJB bulk samples (Amari, Zinner, & Lewis 2000). Hence, the analyzed grains in this study can be considered to be representative of KJB grains. The distribution of $^{12}C/^{13}C$ ratios of the KJB grains (Fig. 1) is compatible with that of micrometer-sized grains, with most grains having $^{12}C/^{13}C$ ratios between 30 and 100. The KJB Y and Z grains have $^{12}C/^{13}C$ ratios of 100-170 and, respectively, 20-80, which is within the ranges of previously identified Y and Z grains.

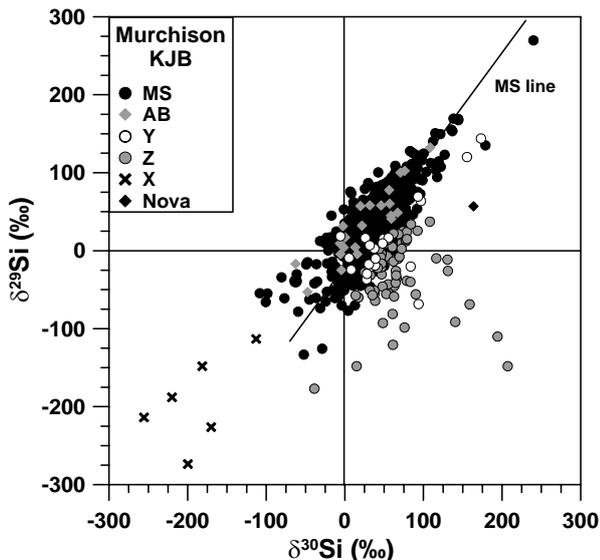

**Figure 2** Si-isotopic compositions of presolar SiC grains from Murchison separate KJB. The different populations are indicated by the different symbols. Several type X grains and the unusual grain KJB2-11-17-1 (Hoppe et al. 2009) are off-scale. The SiC mainstream line ($\delta^{29}Si = 1.37 \times \delta^{30}Si - 20$; Zinner et al. 2007) is shown for reference. Error bars are omitted for clarity; typical errors are <20 ‰ in both directions. $\delta^i Si = \{(^iSi/^{28}Si)_{Grain}/(^iSi/^{28}Si)_\odot - 1\} \times 1000$.

The Si isotope data of the KJB grains are displayed in Fig. 2. In general, the KJB data are similar to those of micrometer-sized grains (Hoppe et al. 1994; Nittler & Alexander 2003), with two exceptions: The clearly higher abundance of Z grains among the KJB grains and the Si-isotopic signature of grain KJB2-11-17-1, which has $\delta^{29}Si = 634 \pm 20$ ‰ and $\delta^{30}Si = -177 \pm 18$ ‰. A Type II supernova origin was considered likely for this grain and its Si-isotopic signature has important implications for the production of $^{29}Si$ in supernovae (Hoppe et al. 2009).

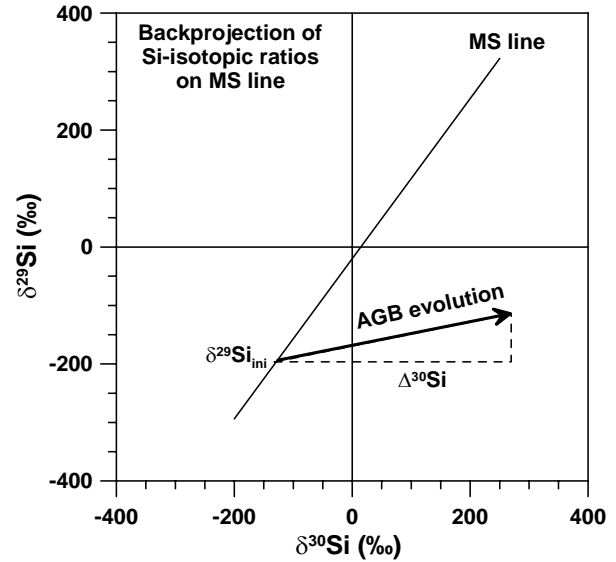

**Figure 3** Backprojection of measured Si-isotopic compositions of Y and Z grains (assumed to plot at the tip of the arrow) along the predicted AGB evolution line on the mainstream (MS) line (Zinner et al. 2006).

Based on their specific C- and Si-isotopic characteristics it was suggested that the Y and Z grains originate from low-mass AGB stars with lower than solar metallicities: 0.5x solar for the Y grain parent stars (Amari et al. 2001) and 0.3x solar for the Z grain parent stars (Hoppe et al. 1997). It is generally accepted that the SiC mainstream line (Zinner et al. 2007) represents essentially the Si starting compositions of the parent stars that contributed dust to the molecular cloud from which our Solar System formed some 4.6 Gy ago. During the AGB phase of stellar evolution the Si isotope abundances are modified by the s-process in the He intershell (Busso et al. 1999). However, in stars of solar metallicity the Si-isotopic compositions of the envelope is affected only marginally by the dredge-up of s-process Si. Only in stars of lower than solar metallicity the imprint of dredge-up of s-process Si becomes clearly visible, which shifts the initial Si-isotopic compositions to the $^{30}Si$-rich side of the SiC mainstream line. It is possible to quantify the effect of s-process Si dredge-up by backprojecting the Si-isotopic ratios of the Y and Z grains along the predicted AGB evolution line on the mainstream line. According to model predictions for AGB stars by Zinner et al. (2006), Si is expected to evolve along a line with slope

~0.2 in a Si-three-isotope delta-plot, if the neutron-capture cross-sections for the Si isotopes from Guber et al. (2003) are used. The intersection between the mainstream line and the AGB evolution line defines two quantities: $\delta^{29}Si_{ini}$ and $\Delta^{30}Si$ (Fig. 3). While the first quantity is a measure for the metallicity of the parent star, the latter is a measure for the s-process Si dredge-up. For the Z grains of this study $\Delta^{30}Si$ values of up to 150 ‰ have been calculated, for the Y grains of up to 60 ‰.

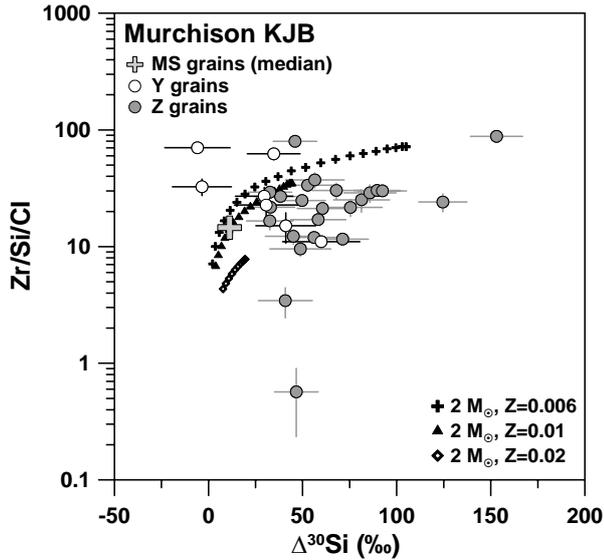

**Figure 4** Zr/Si ratios normalized to solar (CI) in Y and Z grains as a function of $\Delta^{30}Si$ (cf. Fig. 3). Errors are 1σ. Median values for mainstream (MS) grains and model predictions for 2 $M_\odot$ AGB stars with three different metallicities, standard $^{13}C$ pocket, and C/O > 1 are shown for comparison.

The Zr, Mo, Ba, and Nd concentrations were measured in 7 Y and 25 Z grains, and in 23 mainstream grains for comparison. As it turned out, Mo and especially Nd are abundant contaminants in the gold foil and for this reason we focus here on the Zr and Ba data. On average, Zr concentrations are higher in Y and Z grains than in mainstream grains by about a factor of 1.8, which is compatible with expectations for higher s-process contributions in grains from lower than solar metallicity stars. The Ba concentrations in the Z grains are on average higher by about a factor of 2 than in Y and mainstream grains. Normalized to Si and solar (CI), Zr is enriched in Y and Z grains by up to 70x, and Ba by up to 170x. There is a good correlation between Ba concentration and $\Delta^{30}Si$ but not between Zr and $\Delta^{30}Si$ (Figs. 4 and 5).

In the following we compare the Y and Z grain data with predictions for AGB stars by the Torino models, which are based on the FRANEC code (Straniero et al. 1997). Nine different models were investigated by combining M = 1.5, 2, and 3 $M_\odot$ and Z = 0.006, 0.01, and 0.02 (0.3, 0.5, 1 $Z_\odot$), all including the $^{13}C$ neutron source, the main model uncertainty, as in the standard case described by Gallino et al. (1998). Since $\Delta^{30}Si$ is a measure for s-process dredge-up, as are Zr and Ba abundances, one would expect to find correlations between Zr and, respectively, Ba concentrations and $\Delta^{30}Si$, provided these elements were unfractionated incorporated into the growing SiC grains. For Zr there is no match between the grain data and the model predictions (Fig. 4). Most grains have lower Zr concentrations than predicted. This can be explained by removal of Zr from the gas from which the SiC grains condensed, e.g., by condensation of ZrC and/or incorporation of Zr into TiC subgrains, which condense before SiC (Lodders & Fegley 1995) and which have been observed as tiny inclusions in presolar graphite and SiC grains (Bernatowicz et al. 1991; Stroud & Bernatowicz 2005; Croat, Stadermann, & Bernatowicz 2008). Two Y grains have higher Zr concentrations than predicted from unfractionated gas phase condensation and in these cases Zr-rich subgrains might be present. Contamination with terrestrial Zr such as observed for Mo and Nd may be a distinct possibility for these two grains. For Ba the situation is clearly different. There is a positive correlation between Ba abundances and $\Delta^{30}Si$ and most grain data are well matched by the model predictions for lower than solar metallicity 2 and 3 $M_\odot$ AGB stars. This confirms low-metallicity (0.3 – 0.5x solar) AGB stars as most likely stellar sources of these classes of rare presolar SiC grains.

The observed correlation between Ba abundance and $\Delta^{30}Si$ requires largely unfractionated condensation of Ba into the SiC host grains. Condensation calculations (Lodders & Fegley 1995; 1997) do not predict the formation of refractory Ba carbides in AGB star winds but show that Ba may condense into SiC as BaS. Implantation of significant amounts of Ba appears less likely in view of the small size of Z grains, which is smaller than the predicted implantation range (Verchowsky, Wright, & Pillinger 2004).

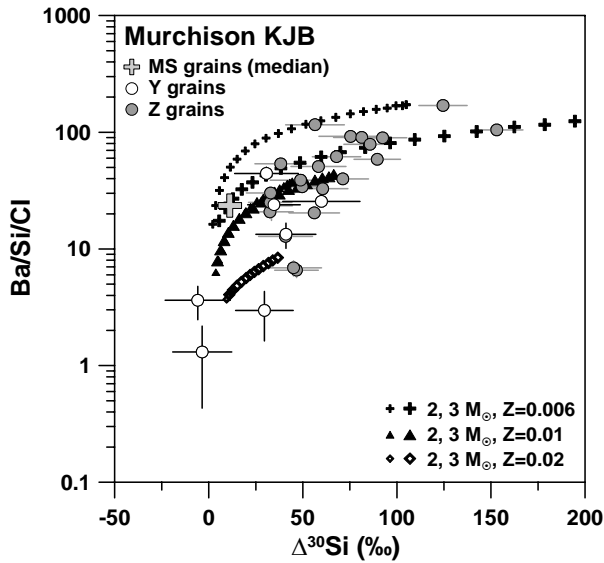

**Figure 5** Ba/Si ratios normalized to solar (CI) in Y and Z grains as a function of $\Delta^{30}$Si (cf. Fig. 3). Errors are $1\sigma$. Medain values for mainstream (MS) grains and model predictions for 2 and 3 $M_\odot$ AGB stars with three different metallicities, standard $^{13}$C pocket, and C/O > 1 are shown for comparison.

## 4. Summary

We have measured C- and Si-isotopic compositions of about 1300 individual, submicrometer-sized SiC grains by NanoSIMS ion imaging. A large number of the rare Y and Z grains could be identified based on their specific C- and/or Si-isotopic signatures. Trace element (Zr, Ba) abundances were studied in selected Y, Z, and mainstream grains. The most striking results are: (1) Zr and Ba concentrations are higher in Y and Z grains than in mainstream grains with enrichments relative to Si and solar of up to 70x (Zr) and 170x (Ba). (2) For the type Y and Z grains there is a positive correlation between Ba concentration and amount of s-process Si. This correlation is well explained by predictions for 2-3 $M_\odot$ AGB stars with metallicities of 0.3-0.5x solar, which confirms low-metallicity stars as most likely stellar sources for the type Y and Z grains.

## Acknowledgements

We thank Maria Lugaro and her team for the organization of the workshop "The origin of the elements heavier than Fe" which was held in Torino in September 2008 and where this paper was presented. We also wish to thank Roy Lewis for the preparation of the KJB mount and Joachim Huth for SEM analyses. We acknowledge the constructive review by an anonymous referee.